# When research assessment exercises leave room for opportunistic behavior by the subjects under evaluation

Giovanni Abramo

*Laboratory for Studies in Research Evaluation*
*at the Institute for System Analysis and Computer Science (IASI-CNR)*
*National Research Council of Italy*
ADDRESS: Istituto di Analisi dei Sistemi e Informatica, Consiglio Nazionale delle Ricerche, Via dei Taurini 19, 00185 Roma - ITALY
giovanni.abramo@uniroma2.it

Ciriaco Andrea D'Angelo

*University of Rome "Tor Vergata" - Italy and*
*Laboratory for Studies in Research Evaluation (IASI-CNR)*
ADDRESS: Dipartimento di Ingegneria dell'Impresa, Università degli Studi di Roma "Tor Vergata", Via del Politecnico 1, 00133 Roma - ITALY
dangelo@dii.uniroma2.it

Flavia Di Costa

*Research Value s.r.l.*
ADDRESS: Research Value, Via Michelangelo Tilli 39, 00156 Roma- ITALY
flavia.dicosta@gmail.com

**Abstract**
This study inserts in the stream of research on the perverse effects that PBRF systems can induce in the subjects evaluated. The authors' opinion is that more often than not, it is the doubtful scientific basis of the evaluation criteria that leave room for opportunistic behaviors. The work examines the 2004-2010 Italian national research assessment (VQR) to verify possible opportunistic behavior by universities in order to limit the penalization of their performance (and funding) due to the presence of scientifically unproductive professors in faculty. In particular, institutions may have favored "gift authorship" practices. The analysis thus focuses on the output of professors who were unproductive in the VQR publication window, but became productive ("new productives") in the following five years: a number of universities show a remarkably higher than average share of publications by new productives that are in co-authorship exclusively with colleagues from the same university.

# 1. Introduction

Governments are clearly aware of the challenges of competitiveness in an increasingly global and knowledge-based economy. Given this, many countries have developed policies for improving the effectiveness and efficiency of their national scientific infrastructure. These have included the introduction of New Public Management (NPM) tools in national research institutions, such as systems of performance-based research funding (PBRF) (Hicks, 2012; Lewis, 2013; Woelert, 2015). PBRF systems were first implemented in the late 1980s and early 1990s in English-speaking countries, and then spread to western Europe, eastern Europe and Asia. By 2016, in the European Union alone, at least nine countries had implemented comprehensive PBRF systems (Jonkers & Zacharewicz, 2016), each with its own national characteristics.

The increasing influence and diffusion of these systems respond to a series of aims, including: i) guaranteeing and improving legitimacy, accountability and effectiveness of public spending in research; ii) increasing awareness of the importance of public research in developing competitive knowledge societies (Whitley & Glaser, 2007; Geuna & Martin, 2003); iii) improving research performance and concentrating resources in the best performing organizations (OECD, 2010; Woelert & McKenzie, 2018; Jonkers & Zacharewicz, 2016; Abramo, 2017).

PBRFs are generally based on national research evaluation systems, which are in turn becoming increasingly common. These national systems can be differentiated according to: i) the choice of units to be evaluated (individuals, departments, entire research institutions); ii) the way in which research activities are evaluated (observation period, type of output evaluated, indicators used, methodology adopted - quantitative vs qualitative forms of evaluations); iii) proportion and types of funding allocation associated with PBRF mechanisms. Government funds are generally allocated at an aggregate level based on the results of the evaluation process, leaving the task of internal allocation to the research institutions.

PBRF systems can have various effects on the strategic and organizational management of the structures under assessment, as well as on the behavior of individual researchers (Geuna & Martin, 2003), particularly when institutions deploy incentives at the individual level, through monetary rewarding systems, access to resources, career advancement, recruitment, etc.. (Moher et al., 2018).

A central question concerning PBRF systems is how and to what extent system-wide incentives can translate into local management practices that influence production effectiveness, efficiency, and recruitment and promotion processes (Espeland & Sauder, 2007; Sauder & Espeland, 2009). Woelert and McKenzie (2018) analyzed the deployment of the Australian national PBRF system within the individual research institutions. The authors "find that universities overwhelmingly replicate the major national PBRF indicators internally. If variation was evident, then [it is] mostly in the form of minor modifications to these indicators, not in the choice of indicators *per se*. Analysis of the Australian case thus demonstrates strong vertical alignment between national and institutional research governance mechanisms as well as considerable convergence in the formal organization and governance of research activities at Australian universities." In Norway, findings are "puzzling" (Aagaard, 2015). In many instances, the author finds "a quite tight coupling between system-level incentives and local practices. A large variation across institutions, fields and departments is, however, also observed."

Some scholars have also considered the important issue of the perverse effects that PBRF systems could have on evaluated subjects, in addition to the expected positive results. These include generation of perverse incentives, inducing scientific misconduct (e.g. multiplication of irrelevant publications, plagiarism, self-plagiarism, scientific fraud) (Hazelkorn, 2010; Edwards & Roy, 2017), and discouraging interdisciplinary and innovative research, and research diversification (Hicks, 2012; Rafols, Leydesdorff, O'Hare, Nightingale, & Stirling, 2012; Wilsdon, 2016, Abramo, D'Angelo & Di Costa, 2018). PBRF systems also present their own direct and indirect costs, which are often underestimated or ignored.

In 2017, the Journal of Informetrics dedicated a special section to PBRF systems and their effects on scientists' behavior (Volume 11, Number 3). The debate opened with a discussion paper by van den Besselaar, Heyman, and Sandström (2017a), who object to the results of Butler's pioneering works (Butler, 2003a; 2003b) on the effects of the Australian PBRF system, in which significant funds were distributed to universities, and then within them, on the basis of aggregate publication counts, with little attention paid to the impact of that output. Butler (2003a; 2003b) found that over the decade examined there had been a 25% increase in Australia's share of publications in the Science Citation Index (SCI), but also a significant decline in the average impact of output. On the contrary, according to van den Besselaar et al. (2017a), the average impact per publication had increased following the introduction of the PBRF.

The paper by van den Besselaar et al. (2017a) was commented by Butler herself (Butler, 2017), and by a number of other experts on performance-based research funding systems (Aagaard & Schneider, 2017; Gläser, 2017; Hicks, 2017; Martin, 2017); van den Besselaar et al. (2017b) then responded. This debate on the Australian case opened to much broader consideration of some fundamental methodological issues in the study of PBRF systems.

Prior to the above-mentioned special section of the Journal of Informetrics, Schneider, Aagaard and Bloch (2016) studied the effects of the PBRF introduced in Norway for the 2006 distribution of university and college funding, extended in subsequent years to health care and public research institutions. The performance evaluation, which examined publication activity, journal publication profiles, and citation impact, used a differentiated point system to discourage researchers from speculating in "easy publications". The *ex post* analysis showed that: overall publication activity increased, impact remained stationary, and activity did not shift to the lowest-impact journals.

The authors' opinion is that more often than not, it is the doubtful scientific basis of the evaluation criteria that leaves room for opportunistic behavior by the subjects under evaluation. The Australian case, stimulus of the above debate, represents just such an instance: if, instead of assessing performance and allocating funding according to number of publications, the allocation were according to their total impact, then there would have been no incentive to produce low-value papers. In fact the ultimate objective of research activity is not to produce papers, but rather scientific advances useful to the scientific community and/or practitioners. The case of the Italian research assessment, VQR 2004-2010, which will be the subject of this study, also left room for opportunistic behaviour, i.e. signing papers without contributing to the research. As we will explain in more detail, the performance evaluation criteria penalized institutions if they submitted lesser numbers of products from their individual professors than the numbers required for the evaluation. Knowing that the full count of products would be evaluated rather than the contributions to these, the institutions could avoid penalization by asking their highly productive researchers to allow unproductive ones to "sign" some of their publications, even where the latter had not made any contribution to the work.

The aim of this study is to verify whether there is any evidence that could be traced back to possible opportunistic behaviour on the part of Italian universities, so as to limit penalization of performance penalties and, therefore, acquisition of funding.

The next section describes the main features of the VQR 2010-2014, which guided allocation of a part of the public funding for Italian universities. Section 3 presents the data and analytical methods, Section 4 the results, and Section 5 the conclusions.

## 2. The 2004-2010 VQR Italian research assessment exercise

Until 2009, the core government funding for Italian universities was input oriented: funds were distributed in a manner that would equally satisfy the needs of each and all, in function of institutional size and disciplines of research. Core funding, known as Ordinary Finance Funds (FFO), accounted for 56% of total university income. It was only following the first national evaluation exercise (2001-2003 VTR) that a minimal share, equivalent to 3.9% of total income, was allocated in function of research and teaching assessments. The launch of the second Italian research assessment exercise (2004-2010 VQR)[1] was preceded by vigorous debate, fueled by heavy cuts in financing to research and higher education under a series of governments. On the one hand were demands that policy makers take courageous action to implement a true PBRF system, capable of attaining improved performance at all levels; in opposition, an insistence on complete renunciation of the assessment program, or at least its serious revision. The VQR thus began in a period of heightened tensions. The purpose of the exercise was to evaluate research activity over the 2004-2010 period as carried out by:
- state universities;
- legally-recognized non-state universities;
- research institutions under the responsibility of the Ministry of Education, University and Research (MIUR).[2]

The MIUR entrusted implementation of the national exercise to the newly formed Agency for the Evaluation of University and Research Systems (ANVUR), which opened the evaluation process on 7 November 2011, and terminated it on 16 July 2013 with the publication of the university performance ranking lists.

The subjects under evaluation were the institutions, their macro-disciplinary areas and departments, but not the individual researchers. The results influence two areas of action: i) overall institutional evaluations have guided allocations of the merit-based share of FFO (13% in 2013, increasing in subsequent years to the current 20%); evaluation of the macro-areas and departments can be used by the universities to guide internal allocation of the acquired resources.

The evaluation of the overall institutions was determined by the weighted sum of a number of indicators: 50% based on a score for the quality of the research products submitted and 50% derived from a composite of six other indicators (10% each for capacity to attract resources, mobility of research staff, internationalization and PhD programs; 5% each for ability to attract research funds and overall improvement since the previous VTR).

---

[1] cfr. the Ministerial decree at http://www.anvur.it/wp-content/uploads/2011/11/vqr_d.m._n._17_del_15_07_2011_firmato.pdf, last accessed 31/10/2018
[2] Other public and private organizations engaged in research could participate in the evaluation by request, subject to fees.

ANVUR appointed 14 evaluation panels (GEVs)[3] of national and foreign experts, one for each university disciplinary area (UDA) in the national academic system. The institutions subject to evaluation were to submit a specific number of products for each person on their research staff, in function of academic rank and their period of activity over the seven years considered. The demand for submissions from university faculty members was up to three products, while for research institutions the maximum was six products per person. ANVUR defined the acceptable products as: a) journal articles; b) books, book chapters and conference proceedings; c) critical reviews, commentaries, book translations; d) patents; e) prototypes, project plans, software, databases, exhibitions, works of art, compositions, and thematic papers.

Any results produced in collaboration with professors in the same institution could only be presented once. Professors were therefore typically asked to identify a set of products larger than the minimal demand, from which the administration could complete the selection of the numbers required for the VQR evaluation. The products were then submitted to the appropriate GEVs based on the professor's identification of the field for each product. The GEVs were to judge the merit of each product as one of four values:

A = Excellent (score 1), if the product places in the top 20% on "a scale of values shared by the international community";
B = Good (score 0.8), if the product places in the 60%-80% range;
C = Acceptable (score 0.5), if the product is in the 50%-60% range;
D = Limited (score 0), if the product is in the bottom 50%.

The institutions are also subject to potential penalties:
  i.  in proven cases of plagiarism or fraud (score -2);
 ii.  for product types not admitted by the GEV, or lack of relevant documentation, or produced outside the 2004-2010 period (score -1);
iii.  for failure to submit the requested number of products (-0.5 for each missing product).

This last penalty considered the nature of the Italian higher education system, which unlike the systems of English-language countries does not provide for both "teaching-only" and research universities. In keeping with the Humboldtian university model, emphasizing the unity of teaching and research, Italy does not have "teaching-only" universities, and all professors are required to carry out both teaching and research.

It should be noted that both the 2004-2010 and subsequent 2011-2014 VQR have received various criticisms concerning the evaluation criteria, performance indicators, distortions in rankings and the resulting allocation of resources (Abramo & D'Angelo, 2017, 2016, 2015, 2013, 2011; Abramo, D'Angelo & Di Costa, 2014; Baccini, 2016; Baccini & De Nicolao, 2016; Franceschini & Maisano, 2017). In this work we explore the possibility that a failure in the evaluation criteria, in addition to inducing distortions in the performance scores and ranks of the institutions, also induces their opportunistic behavior. Leaving room for such behavior does not necessarily imply that it occurs: the extent of the phenomenon very much depends on the cultural traits and core values of the specific system. It suffices to say that Italy shows poor work ethic in the public sector and universities in particular. Academic recruitment and career advancement through centrally regulated competitions have repeatedly come under heavy fire and the term "concorso" has gained currency as a word denoting rigged competition, involving favoritism, nepotism and other unfair selection practices (Gerosa, 2001; Abramo, D'Angelo, & Rosati, 2014; Abramo, D'Angelo, & Rosati, 2015).

---
[3] Acronym of "Groups of Evaluation Experts"

## 3. Data and method

This work aims to verify the presence of any adaptation strategies by Italian universities, adopted for purposes of avoiding the penalties associated with point iii. of the previous section. The 2004-2010 VQR provided for the submission of three research products per professor: their evaluation was independent of the number of co-authors and the position of the submitting author in the byline.[4] Assuming that the criterion remains unchanged in successive VQR exercises, it can be expected that research institutions, to maximize the performance score, will try to induce unproductive professors to produce research outputs. This could occur through serious research work, or simply by "signing" some papers (to which they have not contributed) of more productive and complicit colleagues (who would still have other works to submit to VQR).

The verification that such behaviour could have occurred is conducted by a three-step methodology:
1) We identify the professors on staff in Italian universities in the period 2004-2016 and census their scientific production indexed in the Web of Science (WoS);
2) We identify the subset of professors who were unproductive over the period observed in VQR (2004 to 2010);
3) Assuming that that any adaptation strategies with a view to the subsequent VQR have been undertaken after communication of the 2004-2010 evaluation criteria (which were published in November 2011), we analyze the output of this subset of professors over the 2012-2016 period, tagging any output observed as "new production".

The publication of a given professor is further tagged as:
- "intramural", if resulting from collaborations with authors belonging to his/her university;
- "domestic extramural", if resulting from collaborations with authors belonging to other domestic organizations;
- "international", if resulting from collaborations with authors belonging to foreign organizations.

The verification of the hypothesis is entrusted to analysis of the publication bylines of new productive professors, "new productives", and comparison with the data collected on the rest of the population: if the share of products made only with intramural co-authors (or with internal and external authors) is significantly higher than the average of the entire reference population, it can be assumed that there has been the recourse to adaptation strategies on the part of the universities.

---

[4] On the importance of accounting for the order of authors in the byline, when assessing institutional research performance, see Abramo, D'Angelo and Rosati (2013).

The source for data on the faculty at each university is the database maintained by the Ministry of Education, Universities and Research (MIUR),[5] which indexes the full name, academic rank, research field and institutional affiliation of all professors in Italian universities, at the close of each year. Observed at 31/12/2016, there were 31,381 full, associate and assistant professors working at Italian universities and permanently on staff over the 2004-2016 period. Each is classified in one and only one research field named "scientific disciplinary sector" (SDSs, 370 in all).[6] The SDSs are grouped into disciplines named "university disciplinary areas" (UDAs, 14 in all). To ensure robustness of the bibliometric approach, the dataset is limited to the 20,512 professors in the sciences, in which research output is likely to be indexed in WoS. Table 1 shows their distribution by academic rank and UDA.

The scientific production of such professors is extracted from the Observatory of Public Research (ORP), a database developed by the authors and derived under license from Clarivate Analytics' WoS-Core Collection. Beginning from the raw data of WoS and applying a complex algorithm for disambiguation of the true identity of the authors and reconciliation of their institutional affiliations, each publication is attributed to the university professor that produced it, with a harmonic average of precision and recall (F-measure) equal to 97% (for details see D'Angelo, Giuffrida, & Abramo, 2011).

*Table 1: Dataset of the analysis, by UDA*

| UDA | SDS | Professors | | | |
| --- | --- | --- | --- | --- | --- |
| | | Assistant | Associate | Full | Total |
| Mathematics and computer science | 10 | 376 | 766 | 803 | 1,945 |
| Physics | 8 | 200 | 634 | 476 | 1,310 |
| Chemistry | 12 | 377 | 828 | 566 | 1,771 |
| Earth sciences | 12 | 140 | 292 | 193 | 625 |
| Biology | 19 | 750 | 1,087 | 933 | 2,770 |
| Medicine | 50 | 1,939 | 2,005 | 1,746 | 5,690 |
| Agricultural and veterinary sciences | 30 | 403 | 814 | 704 | 1,921 |
| Civil engineering | 9 | 119 | 381 | 379 | 879 |
| Industrial and information engineering | 42 | 362 | 1,243 | 1,438 | 3,043 |
| Psychology | 8 | 87 | 207 | 264 | 558 |
| Total | 200 | 4,753 | 8,257 | 7,502 | 20,512 |

## 4. Results and analysis

Table 2 presents the descriptive statistics of the unproductive professors ("unproductives") by academic rank in the first and second periods.[7] The percentage of unproductives in the second period, at 5.5% of total, is slightly lower than the corresponding value (5.9%) for the first period. Both of these percentages decrease with increasing academic rank, and for all ranks the percentage of unproductives drops with time: among assistant professors from 13.7% to 12.7%; among associates from 4.8% to 4.5%; among full professors from 2.2% to 2.0%. The balance between new productives and new unproductives is positive in favor of the first for all academic ranks, although this is marginal: the maximum difference is for assistant professors (7.2% vs 6.2%). The always unproductives, i.e. professors who have no publications over both periods, are 2.6% of the total, with a maximum (6.5%) for assistant professors and a minimum for full professors (0.7%).

---

[5] http://cercauniversita.cineca.it/php5/docenti/cerca.php, last accessed 31/10/2018
[6] The complete list is accessible at attiministeriali.miur.it/userfiles/115.htm, last accessed 31/10/2018.
[7] Note that the first period is seven years, compared to the second of five years.

*Table 2: Descriptive statistics of the professors in the different classes, per academic rank*

| Academic rank | Assistant professors | Associate professors | Full professors | Total |
|---|---|---|---|---|
| Observed | 4,753 | 8,257 | 7,502 | 20,512 |
| Unproductives 2004-2010 | 13.7% | 4.8% | 2.2% | 5.9% |
| Unproductives 2012-2016 | 12.7% | 4.5% | 2.0% | 5.5% |
| Always unproductives | 6.5% | 2.0% | 0.7% | 2.6% |
| New productives | 7.2% | 2.8% | 1.4% | 3.3% |
| New unproductives | 6.2% | 2.4% | 1.3% | 2.9% |

Table 3 shows the descriptive statistics for unproductive professors per UDA in the two periods. Initially their presence varies from a minimum of 0.7% in UDA 3 (Chemistry) to a maximum of 24.6% in UDA 11 (Psychology). There are then 676 "new productives" in the 2012-2016 period, i.e. authors of at least one publication indexed in WoS in the period. The distribution per UDA is again obviously uneven: incidence with respect to the unproductives in the first period varies between 25.0% (UDA 2-Physics) and 66.7% (UDA 7-Agricultural and veterinary science). The heterogeneity at UDA level is a reflection of that at the lower level SDS aggregation. In the Mathematics UDA, the share of "new productives" relative to first period unproductives is 38.2%; the statistic varies from a minimum of 0 in the MAT/09 SDS (Operations research) to a maximum of 66.7% in MAT/08 (Numerical analysis). In Civil engineering (UDA 8), incidence of new productives ranges from a minimum (33.3%) in ICAR/01 (Hydraulics) to a maximum (85.7%) in ICAR/04 (Road, railway and airport construction). In Psychology (UDA 11), the overall incidence of new productives is again high (64.2%) and varies between 47.6% in M-PSI/01 (General psychology) and 87.5% in M-PSI/05 (Social psychology). In the seven other UDAs, variation between scientific fields is maximum, since in the second period there is the simultaneous presence of at least one SDS still without new productives, and at least one where all unproductives have become new productives.

*Table 3: Descriptive statistics of the first period unproductives versus new productives in the second period; min-max of new productives in the SDSs of each UDA*

| UDA* | SDSs | Obs | Unproductives in the first period | New productives | Min | | Max | |
|---|---|---|---|---|---|---|---|---|
| 1 | 10 | 1,945 | 144 (7.4%) | 55 (38.2%) | 0.0% | (MAT/09) | 66.7% | (MAT/08) |
| 2 | 8 | 1,310 | 16 (1.2%) | 4 (25.0%) | 0.0% | (various) | 100.0% | (FIS/04) |
| 3 | 12 | 1,771 | 13 (0.7%) | 8 (61.5%) | 0.0% | (CHIM/08) | 100.0% | (various) |
| 4 | 12 | 625 | 34 (5.4%) | 21 (61.8%) | 0.0% | (GEO/06) | 100.0% | (various) |
| 5 | 19 | 2,770 | 58 (2.1%) | 37 (63.8%) | 0.0% | (various) | 100.0% | (various) |
| 6 | 50 | 5,690 | 404 (7.1%) | 213 (52.7%) | 0.0% | (various) | 100.0% | (various) |
| 7 | 30 | 1,921 | 180 (9.4%) | 120 (66.7%) | 0.0% | (various) | 100.0% | (various) |
| 8 | 9 | 879 | 95 (10.8%) | 58 (61.1%) | 33.3% | (ICAR/01) | 85.7% | (ICAR/04) |
| 9 | 42 | 3,043 | 127 (4.2%) | 72 (56.7%) | 0.0% | (various) | 100.0% | (various) |
| 11 | 8 | 558 | 137 (24.6%) | 88 (64.2%) | 47.6% | (M-PSI/01) | 87.5% | (M-PSI/05) |
| Total | 200 | 20,512 | 1,208 (5.9%) | 676 (56.0%) | 0.0% | (various) | 100.0% | (various) |

*First period: 2004-2010; Second period: 2012-2016*

\* 1 - Mathematics and computer science, 2 - Physics, 3 - Chemistry, 4 - Earth sciences, 5 - Biology, 6 - Medicine, 7 - Agricultural and veterinary sciences, 8 - Civil engineering, 9 - Industrial and information engineering, 11 - Psychology

The next step is the verification for evidence of a possible adaptive strategy on the part of the universities, which could have induced/facilitated a process of "conversion" for their first period unproductives. For this, we carry out a comparative analysis of the bylines of publications by new productives in the second period, against those of a control sample of "always productive" professors.

As mentioned, the VQR criteria provided for penalization of universities if they submit a number of products for evaluation per individual professor which is below that required. Since the products presented are evaluated independently of the author's contribution (full counting), the institutions could have favored intramural "guest" authorship, asking highly productive professors to include the names of unproductive ones in the bylines of some of their publications, even if the latter had contributed nothing to the work. The hypothesis could be verified if the share of publications by new productives in collaboration with coauthors from their own organization is observed to be significantly higher than that recorded for the reference population. This is what appears to emerge from the Table 4 data.

*Table 4: Analysis of the bylines of publications authored by new productives and a control sample of always productives; average values and t-test statistics.*

|  | New productives | | Control sample | | | |
|---|---|---|---|---|---|---|
| Observations | 676 | | 990 | | | |
| Authorships | 2,783 | | 13,738 | | | |
|  | Mean | 95% conf. interv. | Mean | 95% conf. interv. | t | |
| Publications per professor | 4.1 | [3.8 - 4.4] | 13.9 | [12.8 - 14.9] | 15.07 | \*\*\* |
| International (%) | 15.3 | [29.9 - 31.4] | 30.6 | [29.9 - 31.4] | 16.56 | \*\*\* |
| Extramural (%) | 58.0 | [56.1 - 59.8] | 57.8 | [57.0 – 58.6] | -0.14 | |
| Intramural (%) | 80.3 | [78.8 - 81.8] | 78.0 | [77.3 - 78.7] | -2.67 | \*\*\* |
| Exclusively intramural (%) | 32.8 | [31.1 - 34.6] | 25.7 | [25.0 - 26.4] | -70.7 | \*\*\* |
| Single author (%) | 2.6 | [2.0 - 3.1] | 1.5 | [1.3 - 1.7] | -4.15 | \*\*\* |
| No. of co-authors | 6.3 | [5.7 - 6.9] | 8.3 | [7.7 - 8.9] | 2.87 | \*\*\* |
| No. of co-authors, excluding mega-authorships† | 5.9 | [5.7 - 6.0] | 6.6 | [6.5 - 6.7] | 6.32 | \*\*\* |

† *publications with 100 co-authors or more*
\* p <0.1; \*\* p < 0.05; \*\*\* p <0 .01

As expected, the individual output of new productives (4.1 publications per professor) is significantly lower than that of the "always productives" (13.9). Turning to data on their collaborative publications, we observe that the share of publications produced with foreign colleagues is, for new productives, half of that recorded for the always productives: 15.3% vs 30.6%. The share of publications resulting specifically from extramural domestic collaborations is similar for the two sets of professors; the difference between averages (58.0% vs 57.8%) is not statistically significant. On the contrary, intramural coauthorships are more frequent for new productives, with statistically significant difference, at about two percentage points (80.3% vs 78.0%). Observing the publications resulting from exclusively intramural collaborations, the difference between the two sets grows: 32.8% of the new productives' publications show a byline composed of all and only colleagues from the same university; for "always productives" the statistic is seven points less, at 25.7%. Finally, the bylines for new productives' publications are shorter: on average 6.3 coauthors per publication for new productives, compared to 8.3 for always productives; excluding mega-author publications, the difference between subpopulations is less (5.9 vs 6.6).

These results require a series of reflections and hypotheses concerning the underlying phenomena. Focusing on the publications in intramural (including some extramural) and exclusively intramural coauthorship, the significant differences between new productives and always productives would suggest the possibility of an opportunistic behavior by universities, for purposes of reducing penalties. However it must also be considered that the long-standing unproductives will find it easier to return to publishing by establishing collaborations within their home university, for reasons related to proximity and organizational culture. It is certainly more difficult to establish new collaborations with foreign partners, unless one has a continuing presence in consolidated networks (e.g. PhD obtained abroad, participation in supranational projects awarded in competition): situations difficult to associate with the status of "scientifically unproductive". The domestic extramural collaborations have similar incidences for the two groups compared, so do not seem to be a "discriminating" factor for the hypothesis verification.

The data on average number of co-authors (observed to be less for new productives) could be interpreted in two ways: while the presence of a "spurious" author might seem more tolerable in a large team, a smaller team could accept such a presence if there were strong links between its members, as in the case of colleagues in the same group/department.

Finally, it should be noted that the share of single-author papers is 2.6% of total for new productives compared to 1.5% for the always productives, suggesting that there indeed professors who have reacted virtuously to the stimuli for improvement embedded in the evaluation processes. The phenomenon of "new" single authorship could be predominant in scientific fields of a mainly theoretical nature and less so in those of an experimental nature, where the need for facilities and instrumentation demands cooperative work. In detail, 54 (8% of the total) new productives have at least one single-author paper to their credit, compared to 11.9% for the benchmark. Seven of these (five in the Mathematics and computer science area) have two single-author publications, three have authored 3, and one even 5.

Having seen that the above examination of the data fails to reject the opportunistic behavior hypothesis, we now analyze the incidence of new productives in the unproductives of the first period, for each university (Table 5), to identify institutions that have been more effective in converting unproductive professors into productive. It will also be interesting to measure the share of intramural and exclusively intramural works of the new productives per individual university (Table 6), recalling that the averages for the always productives are respectively 78.0% and 25.7%.

Table 5, column 5 shows the ratio between the percentage of new productives and percentage of first period unproductives.[8] This is a measure of the effectiveness of the incentive systems used by the universities to induce publication by their unproductives. Only the University of Tuscia is fully effective (100% conversion), followed by Reggio Calabria 'Mediterranean' (92.9%) and Insubria-Varese (85.7%). In total there are 31 universities (out of 48 with at least 100 professors in the dataset) with a conversion rate

---

[8] For reasons of significance, the analysis considers only universities with at least 100 professors in the dataset.

higher than 50%. Only Rome 'Tre' does not record any new productives in 2012-2016 relative to the 1.9% of unproductives in 2004-2010.

If, on the other hand, we analyze the scientific production of new productives, and in particular the percentage of publications produced in collaboration with colleagues from the same institution (Table 5, columns 6-8), the University of Udine tops the national ranking. Over 2013-2016 its four new productives achieved 31 publications, of which 29 (93.5% of the total) in intramural co-authorship and 77.4% in exclusively intramural co-authorship. The University of Calabria also recorded a 77.4% share of exclusively intramural coauthorship publications, followed by Genoa (62.5%), Modena-Reggio Emilia (48.4%). These universities represent cases of nearly double to nearly quadruple the rate of exclusively intramural coauthorship publications in the sample of always productives, observed at 25.7%. Interestingly, there is no geographical characterization of the upper or lower part of this ranking. As the funnel plot in Figure 1 shows, although the differences between universities are not always statistically significant, eight universities are above the 3-SD band.[9]

These two analyses therefore confirm that universities have decisively mobilized following the outcome of the first VQR in an attempt to encourage the conversion of unproductives. Where the percentages of publications in collaboration with new productives are significantly higher than the average of those of the always productive, the suspicion of possible opportunistic behaviour is legitimate.

---

[9] The funnel plot is generally used to illustrate how the size of a university affects the variability of the plotted indicator. For details see Abramo, D'Angelo and Grilli (2015).

*Table 5: Percentage of "new productive" publications produced in intramural coauthorship, per university*

| University | Professors | Unproductives in first period (%) - a | New productives (%) – b | Conversion rate (%) – b/a | Publications by new productives | Of which intramural (%) (benchmark 78.0%) | Of which exclusively intramural (%) (benchmark 25.7%) |
|---|---|---|---|---|---|---|---|
| Udine | 265 | 4.2 | 1.5 | 36.4 | 31 | 93.5 | 77.4 |
| Calabria | 253 | 5.5 | 2.4 | 42.9 | 31 | 83.9 | 77.4 |
| Genoa | 501 | 7.0 | 2.2 | 31.4 | 32 | 87.5 | 62.5 |
| Modena-Reggio Emilia | 333 | 3.6 | 1.8 | 50.0 | 31 | 87.1 | 48.4 |
| Ancona Polytechnic | 250 | 2.8 | 1.6 | 57.1 | 25 | 84.0 | 48.0 |
| Turin Polytechnic | 371 | 5.1 | 2.7 | 52.6 | 38 | 84.2 | 47.4 |
| Milan Polytechnic | 531 | 6.6 | 2.8 | 42.9 | 91 | 78.0 | 46.2 |
| Basilicata | 173 | 12.1 | 7.5 | 61.9 | 52 | 73.1 | 46.2 |
| Palermo | 549 | 8.0 | 5.3 | 65.9 | 148 | 94.6 | 45.9 |
| Reggio Calabria 'Mediterranean' | 104 | 13.5 | 12.5 | 92.9 | 48 | 91.7 | 45.8 |
| Padua | 837 | 3.3 | 1.9 | 57.1 | 50 | 90.0 | 44.0 |
| Gabriele D' Annunzio | 214 | 7.0 | 4.2 | 60.0 | 19 | 84.2 | 42.1 |
| Salento | 180 | 5.0 | 2.8 | 55.6 | 17 | 88.2 | 41.2 |
| Pisa | 653 | 2.6 | 1.1 | 41.2 | 22 | 77.3 | 40.9 |
| Bologna | 1021 | 3.5 | 1.6 | 44.4 | 73 | 91.8 | 39.7 |
| Turin | 660 | 5.8 | 3.9 | 68.4 | 70 | 75.7 | 38.6 |
| Pavia | 415 | 4.8 | 2.7 | 55.0 | 40 | 75.0 | 37.5 |
| Messina | 479 | 10.0 | 6.1 | 60.4 | 113 | 81.4 | 37.2 |
| Cagliari | 387 | 7.2 | 4.1 | 57.1 | 63 | 76.2 | 36.5 |
| Bari Polytechnic | 154 | 7.8 | 5.8 | 75.0 | 44 | 90.9 | 34.1 |
| Naples 'Second' | 397 | 7.1 | 4.3 | 60.7 | 56 | 69.6 | 33.9 |
| Verona | 184 | 2.7 | 2.2 | 80.0 | 18 | 88.9 | 33.3 |
| Trento | 153 | 3.9 | 2.6 | 66.7 | 9 | 55.6 | 33.3 |
| Siena | 286 | 5.6 | 1.7 | 31.3 | 12 | 75.0 | 33.3 |
| Naples 'Federico II' | 1,108 | 5.6 | 3.5 | 62.9 | 185 | 85.4 | 32.4 |
| dell'Aquila | 324 | 6.2 | 2.2 | 35.0 | 30 | 83.3 | 30.0 |
| Rome 'La Sapienza' | 1,710 | 8.8 | 5.1 | 58.7 | 447 | 86.6 | 29.3 |
| Milan Bicocca | 249 | 6.8 | 4.8 | 70.6 | 42 | 76.2 | 28.6 |
| Catholic Sacred Heart | 528 | 7.2 | 2.7 | 36.8 | 50 | 78.0 | 28.0 |
| Florence | 712 | 6.5 | 3.1 | 47.8 | 91 | 80.2 | 27.5 |
| Ferrara | 280 | 4.6 | 2.9 | 61.5 | 41 | 80.5 | 26.8 |
| Insubria-Varese | 164 | 4.3 | 3.7 | 85.7 | 56 | 60.7 | 25.0 |
| Trieste | 303 | 8.9 | 5.3 | 59.3 | 32 | 53.1 | 25.0 |
| Brescia | 223 | 5.4 | 3.6 | 66.7 | 34 | 64.7 | 23.5 |
| Perugia | 400 | 6.0 | 5.0 | 83.3 | 97 | 81.4 | 21.6 |
| Salerno | 281 | 2.5 | 1.8 | 71.4 | 21 | 47.6 | 19.0 |
| Catania | 568 | 7.4 | 4.2 | 57.1 | 76 | 92.1 | 15.8 |
| Milan | 863 | 3.4 | 1.6 | 48.3 | 72 | 81.9 | 15.3 |
| Parma | 444 | 4.3 | 2.3 | 52.6 | 20 | 60.0 | 15.0 |
| Bari | 581 | 5.5 | 2.4 | 43.8 | 43 | 53.5 | 14.0 |
| Rome 'Tor Vergata' | 633 | 5.2 | 2.5 | 48.5 | 54 | 55.6 | 11.1 |
| Camerino | 130 | 4.6 | 3.1 | 66.7 | 19 | 78.9 | 10.5 |
| Sassari | 238 | 8.8 | 5.0 | 57.1 | 22 | 81.8 | 9.1 |
| Tuscia | 109 | 7.3 | 7.3 | 100.0 | 32 | 68.8 | 3.1 |
| Piedmont 'Orientale A. Avogadro' | 119 | 1.7 | 0.8 | 50.0 | 2 | 100.0 | 0.0 |
| Urbino 'Carlo Bo' | 122 | 9.8 | 2.5 | 25.0 | 10 | 70.0 | 0.0 |
| Rome 'Tre' | 154 | 1.9 | 0.0 | 0.0 | n.a. | - | - |
| Total | 20,512 | 5.9 | 3.3 | 56.0 | 2783 | 80.3 | 32.8 |

*Figure 1: Funnel plot of the share of exclusively intramural publications authored by "new productives" in Italian universities (University of Rome 'La Sapienza' omitted because off scale)*

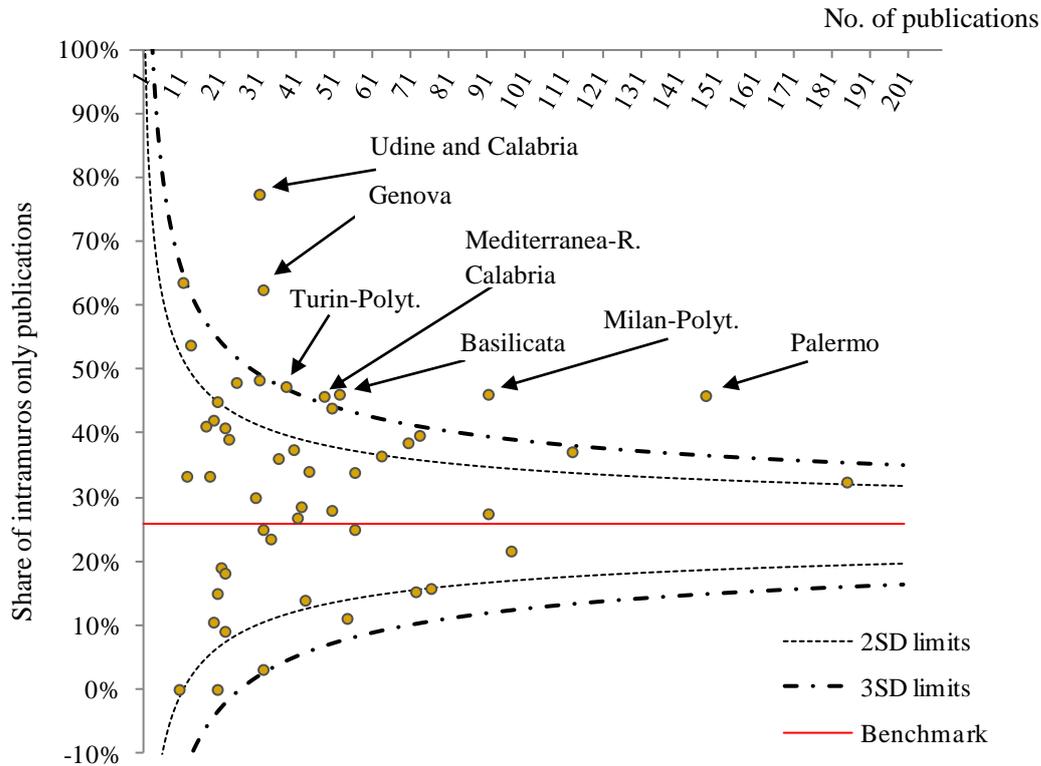

## 5. Conclusions

This paper examines the 2004-2010 Italian research assessment (VQR), to verify for the possible onset of opportunistic behavior on the part of universities, following publication of the procedural criteria, to avoid penalization of performance and consequent reductions in funding due to unproductive professors. Opportunistic behavior is expressed in giving the opportunity to unproductive individuals to sign works in which they have not participated. To this end, we have analyzed the output produced in the 2012-2016 period by professors who were unproductive through the years of the preceding VQR. In particular, a comparative analysis of the bylines of such output was carried out with respect to the scientific production of a control sample of professors, who were always productive over both observed periods.

The results obtained from the analysis show first of all that the individual output of the "new productives" is on average much lower than that of the "always productives". Moreover, the bylines of the new productives' publications demonstrate a high incidence of intramural co-authorship, while their international authorships are about half of the value found for the "always productives". There are no differences for domestic extramural collaborations.

As expected, a scientist who has not published in 7 years can begin to do so but their productivity is likely to be more modest than that of the always productives, especially for publications in international collaboration. Moreover, it will be easier for new productives to begin scientific activity with the support of colleagues who are institutionally closer. Some new productives also show that they can "do it themselves", as demonstrated by the data on the incidence of single-author papers, on average seen higher for new productives than found on the control sample. However, in general, some of the findings make it legitimate to suspect opportunistic behaviour on the part of universities. In particular, some of them show a share of intra-mural publications by new productives significantly above the reference sample. The possibility that some very productive professors allowed their works to be signed by unproductive colleagues, to avoid penalties to the university, cannot be excluded

This empirical evidence should be deepened through specific analyses which, given the delicacy of the subject, may not be easy to conduct. An immediate important observation is that greater caution in formulating the performance evaluation criteria (the adoption of fractional counting of products, for example) could certainly avoid occurrence of the opportunistic behaviour.

The theme of the effect of incentives on academic performance represents a pragmatic, consequential field of research. The authors intend to continue to deepen the subject by shifting the focus to the effects of the evaluation process known as the "National Scientific Habilitation" on the intensity of scientific activity, again by Italian academics.